\documentclass[12pt]{article}
\usepackage{epsfig}
\def\be{\begin{equation}}
\def\ee{\end{equation}}
\def\bea{\begin{eqnarray}}
\def\eea{\end{eqnarray}}
\usepackage{graphicx}

\catcode`\@=11
\def\lsim{\mathrel{\mathpalette\@versim<}}
\def\gsim{\mathrel{\mathpalette\@versim>}}
\def\@versim#1#2{\vcenter{\offinterlineskip
\ialign{$\m@th#1\hfil##\hfil$\crcr#2\crcr\sim\crcr } }}
\catcode`\@=12

\catcode`@=12
\begin{document}
\thispagestyle{empty}
\begin{flushright}
UCRHEP-T549\\
December 2014\
\end{flushright}
\vspace{0.6in}
\begin{center}
{\LARGE \bf Syndetic Model of Fundamental Interactions\\}
\vspace{1.8in}
{\bf Ernest Ma\\}
\vspace{0.2in}
{\sl Physics \& Astronomy Department and Graduate Division,\\ 
University of California, Riverside, California 92521, USA\\}
\end{center}
\vspace{1.8in}

\begin{abstract}\
The standard model of quarks and leptons is extended to connect three 
outstanding issues in particle physics and astrophysics: (1) the absence 
of strong $CP$ nonconservation, (2) the existence of dark matter, and (3) 
the mechanism of nonzero neutrino masses, and that of the first family of 
quarks and leptons, all in the context of having only one Higgs boson in 
a renormalizable theory.  Some phenomenological implications are discussed.
\end{abstract}

\newpage
\baselineskip 24pt

With the 2012 discovery~\cite{atlas12,cms12} of the 125 GeV particle at 
the Large Hadron Collider (LHC), and the likelihood of it being the one 
physical neutral Higgs boson $h$ of the standard model (SM) of quarks and 
leptons, it appears that the SM is essentially complete.  Nevertheless, 
there are at least three loose ends. (1) The SM predicts a source of strong 
$CP$ nonconservation (the neutron electric dipole moment) which is not 
observed. (2) The SM does not have a suitable candidate for the dark 
matter (DM) of the Universe. (3) The SM does not specify a unique fundamental 
(renormalizable) mechanism for nonzero neutrino masses, which are 
required for neutrino oscillations.
Whereas well-known piecemeal solutions of each problem exist, it is worth 
exploring the possibility that these three issues are in fact interconnected. 
I propose in the following a syndetic model with this idea in mind.\\[5pt]

\begin{table}[htb]
\begin{center}
\begin{tabular}{|c|c|c|}
\hline
field & $SU(3) \times SU(2) \times U(1)$ & $U(1)_{PQ}$ \\ 
\hline \hline
$(t,b)_L,(c,s)_L$ & (3,2,1/6) & 0 \\ 
$(u,d)_L$ & (3,2,1/6) & 2 \\ 
$t_R,c_R,u_R$ & (3,1,2/3) & 0 \\ 
$b_R,s_R,d_R$ & (3,1,--1/3) & 0 \\ 
\hline
$(\nu_\tau,\tau)_L,(\nu_\mu,\mu)_L,(\nu_e,e)_L$ & (1,2,--1/2) & 0 \\ 
$\tau_R,\mu_R$ & (1,1,--1) & 0 \\ 
$e_R$ & (1,2,--1/2) & --2 \\ 
\hline
$(\phi^+,\phi^0)$ & (1,2,1/2) & 0 \\ 
\hline \hline
$Q_L$ & (3,1,--1/3) & --1 \\ 
$Q_R$ & (3,1,--1/3) & 1 \\ 
$N_{1R},N_{2R},N_{3R}$ & (1,1,0) & 1 \\ 
\hline
$(\eta^+,\eta^0)$ & (1,2,1/2) & 1 \\ 
$\chi^+$ & (1,1,1) & 1 \\  
$\chi^0$ & (1,1,0) & 1 \\ 
$\zeta^0$ & (1,1,0) & 2 \\
\hline \hline
\end{tabular}
\end{center}
\caption{Field content of proposed model.}
\end{table}

Under the proposed well-known anomalous $U(1)_{PQ}$ symmetry~\cite{pq}, the 
SM fermions transform as shown in Table 1.  The new particles all have 
$PQ$ charges.  They include a heavy singlet quark $Q$ of charge ${-1/3}$ 
and three heavy neutral singlet fermions $N_{1R},N_{2R},N_{3R}$, as well as 
one scalar doublet $(\eta^+,\eta^0)$, one scalar singlet $\chi^+$ and one 
scalar singlet $\chi^0$.  Another scalar singlet $\zeta^0$ with two units 
of $PQ$ charge contains the axion~\cite{we78,wi78} as $U(1)_{PQ}$ is 
spontaneously broken.  This solves the strong $CP$ problem.   A residual 
discrete $Z_2$ symmetry also exists in this case, and acts as a good symmetry 
for cold dark matter, as was pointed out recently~\cite{dmt14}.  Since some 
of the SM quarks as well as the heavy $Q$ transform under $U(1)_{PQ}$, this 
model is a hybrid of the two well-known examples of realistic axion models: 
one where only $Q$ transforms~\cite{k79,svz80} (KSVZ) and one where there is 
no $Q$ but the SM quarks transform~\cite{dfs81,z80} (DFSZ).  The important 
distinction in the present proposal is that the SM Higgs doublet couples at 
tree level only to the second and third families of quarks and charged 
leptons, i.e. they do not transform under $U(1)_{PQ}$.   On the other hand, 
the first family becomes massive only through interactions with $Q$ and $N$ 
in the presence of $U(1)_{PQ}$ symmetry, thus linking their 
origin of mass through the one Higgs boson $h$ of the SM with dark 
matter~\cite{m14} as well as the solution of the strong $CP$ 
problem~\cite{dmt14}.  Because there is only one $Q$ which couples to 
$\zeta$ in Ref.~\cite{dmt14} as in the KSVZ model, 
the domain wall number is 1 in those cases, and that model is 
cosmologically safe~\cite{s82}.   Here the axion couples to both $Q$ as 
well as $u$ and $d$, so it suffers the same domain wall problem as the DFSZ 
model which requires new physics interventions~\cite{s82}.  As for neutrinos, 
they will acquire nonzero Majorana masses, using the {\it scotogenic} 
mechanism~\cite{m06}, from the Greek {\it scotos} meaning darkness.

Under the exactly conserved discrete $Z_2$ symmetry, all SM particles are 
even.  The dark sector consists of particles odd under $Z_2$, namely $Q$, 
$N_1$, $N_2$, $N_3$, $(\eta^+,\eta^0)$, $\chi^+$, and $\chi^0$.  Their only 
direct interactions with $\zeta^0$ are
\begin{equation}
f_Q \zeta^0 \bar{Q}_R Q_L, ~~~ f_N^{ij} \bar{\zeta}^0 N_{iR} N_{jR}, ~~~ 
\lambda_\chi \bar{\zeta}^0 
\chi^0 \chi^0.
\end{equation}
Thus $Q$ and $N$ are expected to acquire large masses from the vacuum 
expectation value of $\zeta^0$, whereas the mass-squared matrix spanning 
$(\chi_R, \chi_I)$ where $\chi^0 = (\chi_R + i \chi_I)/\sqrt{2}$ is of the form
\begin{equation}
{\cal M}^2_\chi = \pmatrix{m_\chi^2 + \mu_\chi \langle \zeta^0 \rangle & 0 
\cr 0 & m_\chi^2 - \mu_\chi \langle \zeta^0 \rangle}.
\end{equation}
This is suggestive of having $\chi_I$ as the lightest particle of odd $Z_2$ 
and thus a DM candidate.  

Since $(u,d)_L$ has PQ charge 2, it does not couple to $\Phi$.  Hence 
$u$ and $d$ quarks are massless at tree level.  However, the mixing of 
$\eta^+$ with $\chi^+$ and $\eta^0$ with $\chi^0$ through $\langle \phi^0 
\rangle$ allows both to acquire small masses in one loop as shown.
\begin{figure}[htb]
\vspace*{-3cm}
\hspace*{-3cm}
\includegraphics[scale=1.0]{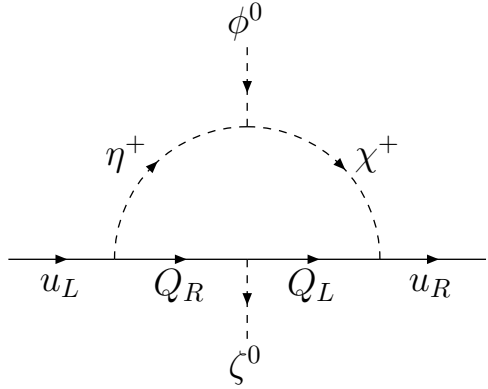}
\vspace*{-20.5cm}
\caption{One-loop generation of $u$ quark mass.}
\end{figure}
\begin{figure}[htb]
\vspace*{-3cm}
\hspace*{-3cm}
\includegraphics[scale=1.0]{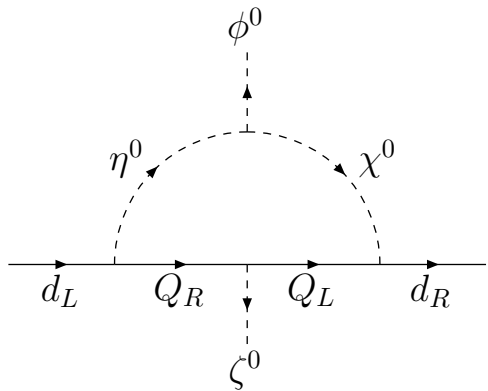}
\vspace*{-20.5cm}
\caption{One-loop generation of $d$ quark mass.}
\end{figure}

The resulting $3 \times 3$ mass matrix linking $\bar{q}_L$ to $q_R$ for either 
sector is of the form
\begin{equation}
{\cal M}_q = \pmatrix{m_{11} & m_{12} & m_{13} \cr 0 & m_{22} & m_{23} \cr 
0 & 0 & m_{33}},
\end{equation}
where $m_{33}, m_{22}, m_{23}$ are tree-level, and $m_{11}, m_{12}, m_{13}$ 
are one-loop radiative masses.  Since the mass eigenvalues in each quark 
sector are hierarchical, and the (13) and (23) mixing angles are very small, 
each mass matrix is diagonalized on the left by
\begin{equation}
U = \pmatrix{1 & 0 & -\epsilon_{13} \cr 0 & 1 & -\epsilon_{23} \cr 
\epsilon_{13} & \epsilon_{23} & 1} \pmatrix{\cos \theta & -\sin \theta & 0 \cr 
\sin \theta & \cos \theta & 0 \cr 0 & 0 & 1},
\end{equation}
where $\epsilon_{13} = m_{13}/m_{33}$, $\epsilon_{23} = m_{23}/m_{33}$, and 
$\tan \theta = m_{12}/m_{22}$.  The mass eigenvalues are then
\begin{equation}
m_t (m_b) \simeq m_{33}, ~~~ m_c (m_s) \simeq \sqrt{m_{22}^2 + m_{12}^2}, 
~~~ m_u (m_d) \simeq m_{11} \cos \theta,
\end{equation} 
and the quark mixing matrix $U_u^\dagger U_d$ has the elements
\begin{eqnarray}
V_{us} &\simeq& \sin (\theta_u - \theta_d), \\ 
V_{ub} &\simeq& \cos \theta_u (\epsilon^u_{13} - \epsilon^d_{13}) 
+ \sin \theta_u (\epsilon^u_{23} - \epsilon^d_{23}), \\ 
V_{cb} &\simeq& \cos \theta_u (\epsilon^u_{23} - \epsilon^d_{23}) 
- \sin \theta_u (\epsilon^u_{13} - \epsilon^d_{13}). 
\end{eqnarray}
Numericaly, in the $d$ quark sector, the radiative entries 
$m_{11} \simeq 5$ MeV and $m_{12} \simeq 22$ MeV, assuming $\theta_u$ is 
very small, whereas the tree-level mass $m_{22} \simeq 92$ MeV. 
This requires the tree-level $m_s$ to be small compared to the tree-level 
$m_b$, but not very large compared to the radiative mass terms.  In other 
words, there is no theoretical understanding of these mass ratios in this 
model.

Radiative electron and Majorana neutrino masses are also generated.
\begin{figure}[htb]
\vspace*{-3cm}
\hspace*{-3cm}
\includegraphics[scale=1.0]{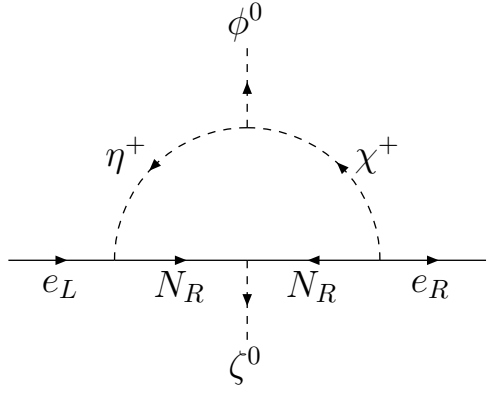}
\vspace*{-20.5cm}
\caption{One-loop generation of electron mass.}
\end{figure}
\begin{figure}[htb]
\vspace*{-3cm}
\hspace*{-3cm}
\includegraphics[scale=1.0]{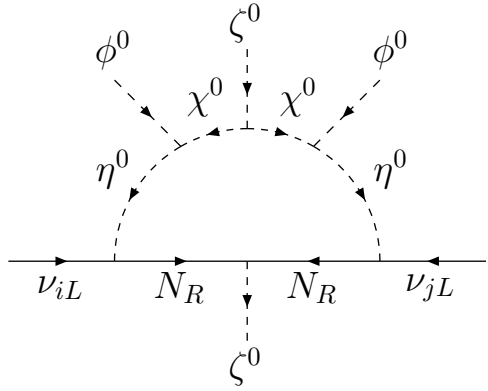}
\vspace*{-20.5cm}
\caption{One-loop generation of Majorana neutrino mass.}
\end{figure}
In contrast to Eq.~(3), the resulting $3 \times 3$ charged-lepton mass 
matrix linking $\bar{l}_L$ to $l_R$ is of the form
\begin{equation}
{\cal M}_l = \pmatrix{m_{11} & 0 & 0 \cr m_{21} & m_{22} & 0 \cr 
m_{31} & m_{32} & m_{33}},
\end{equation}
where $m_{33}, m_{22}, m_{32}$ are tree-level, and $m_{11}, m_{21}, m_{31}$ 
are one-loop radiative masses.  As a result, it is diagonalized on the left by
\begin{equation}
U = \pmatrix{1 & -(m_e/m_\mu)\epsilon_{21} & -(m_e/m_\tau)\epsilon_{31} \cr 
(m_e/m_\mu)\epsilon_{21} & 1 & -(m_\mu/m_\tau)\epsilon_{32} \cr 
(m_e/m_\tau)\epsilon_{31} & (m_\mu/m_\tau)\epsilon_{32} & 1}, 
\end{equation}
where $\epsilon_{21} = m_{21}/m_{22}$, $\epsilon_{31} = m_{31}/m_{33}$, 
$\epsilon_{32} = m_{32}/m_{33}$, and 
$m_\tau \simeq m_{33}$, $m_\mu \simeq m_{22}$, $m_e \simeq m_{11}$.
As for the $3 \times 3$ neutrino mass matrix, it is all radiative as in 
the original scotogenic model~\cite{m06}.  Without the imposition of a 
flavor symmetry, there are no specific predictions of mixing angles or 
masses.

The radiative mass entries of Eqs.~(3) and (9) are of the form~\cite{m14,fm14}
\begin{equation}
m = {f_\eta f_\chi \mu v \over 16 \pi^2 \sqrt{2} M (x_1-x_2)} 
\left( {x_1 \ln x_1 \over x_1 -1} - {x_2 \ln x_2 \over x_2 -1} \right),
\end{equation}
where $f_\eta$ and $f_\chi$ are the respective Yukawa couplings in Figs.~1, 2, 
and 3, $\mu$ is the corresponding trilinear scalar coupling to $\phi^0$, 
$v/\sqrt{2} = 174$ GeV is the vacuum expectation value of $\phi^0$, $M$ 
is either the mass of $Q$ or $N$, $x_{1,2} = m_{1,2}^2/M^2$, and $m^2_{1,2}$ 
are the eigenvalues of the $(\eta,\chi)$ mass-squared 
matrix.  Note that $M$ comes from the vacuum expectation value of $\zeta$.

If all fermion masses in a given sector, say $(d,s,b)$ come from 
tree-level couplings due to a single Higgs boson, then the diagonalization 
of that mass matrix automatically diagonalizes the Higgs Yukawa couplings.  
However, since some of the mass entries are one-loop effects, this will 
not be the case, because the corresponding Higgs Yukawa coupling is then 
not exactly 
equal to $m/v$, as pointed out recently~\cite{fm14}.  As a result, there 
will be small off-diagonal Higgs Yukawa couplings to quarks, which 
induces small flavor-changing neutral-current processes.  The most 
sensitive probe of this effect is $K^0 - \bar{K}^0$ mixing in the quark 
sector and $\mu \to e \gamma$ in the lepton sector.

Since $m_{11}$, $m_{12}$, and $m_{13}$ in Eq.~(3) are radiative masses, the 
corresponding Higgs Yukawa coupling matrix is of the form
\begin{equation}
{\cal H}_q = {1 \over v} \pmatrix{m_{11} (1 + \delta) & m_{12} (1 + \delta) & 
m_{13} (1 + \delta) \cr 0 & m_{22} & m_{23} \cr 0 & 0 & m_{33}}, 
\end{equation}
where $\delta$ is a loop factor computed exactly in Ref.~\cite{fm14}.  Let 
$U_L {\cal M}_q U_R^\dagger$ be diagonal, then $U_L {\cal H}_q U_R^\dagger$ 
will have off-diagonal pieces, i.e. flavor-changing neutral currents. 
Using Eq.~(4) and the fact that $U_R$ has suppressed off-diagonal entries 
relative to $U_L$, the dominant effective operator for $K^0 - \bar{K}^0$ 
mixing through Higgs exchange is given by
\begin{equation}
{\cal O}_2 = {\delta^2 m_s^2 \sin^2 \theta \cos^2 \theta \over v^2 m_h^2} 
(\bar{d}_L s_R)^2.
\end{equation}
Using $v = 246$ GeV, $m_h = 125$ GeV, and $m_s = 55$ MeV, this contribution 
to the $K_L - K_S$ mass difference is $-1.9 \times 10^{-14} \delta^2$ 
GeV~\cite{bm14}, as compared to the experimental value of 
$3.484 \times 10^{-15}$ GeV.  Allowing for a 10\% uncertainty in the 
SM contribution, this means that $\delta$ may be as large as 0.135.

In the charged-lepton sector, after diagonalizing Eq.~(9), flavor violating 
decays such as $\mu \to e \gamma$ will occur.  This is dominated by the 
radiative transition $l_{2L} \to l_{1R} \gamma$ and its amplitude is 
proportional to $m_{21} \sim \epsilon_{21} m_\mu$.  Its calulation is 
analogous to that of the muon anomalous magnetic moment given in 
Ref.~\cite{fm14}, i.e.
\begin{equation}
{\cal A} = {m_{21} \over 2 m_\mu m_N^2} \left[ {G(x_1) - G(x_2) \over H(x_1) 
- H(x_2)} \right],
\end{equation}
where
\begin{equation}
G(x) = {2 x \ln x \over (x-1)^3} - {x+1 \over (x-1)^2}, ~~~ 
H(x) = {x \ln x \over x-1}.
\end{equation}
The branching fraction of $\mu \to e \gamma$ is constrained by the 
current experimental upper bound~\cite{meg13} according to
\begin{equation}
B = {12 \pi^3 \alpha \epsilon_{21}^2 \over G_F^2 m_N^4} \left[ {G(x_1) 
- G(x_2) \over H(x_1) - H(x_2)} \right]^2 < 5.7 \times 10^{-13}.
\end{equation}
For $m_N < 1$ TeV, this requires $\epsilon_{21}$ to be less than about 
$10^{-5}$.  This implies that a flavor symmetry, e.g. $Z_3$, is desirable 
in a more complete model to make $m_{21}$ zero.

At the Large Hadron Collider, the heavy quark $Q$ may be produced in 
pairs if kinematically allowed.  Consider the decay chain:
\begin{equation}
Q^{-1/3} \to u + \eta^-, ~~~ \eta^- \to e^- N ~{\rm or}~ \mu^- N,
\end{equation}
then if $N$ is dark matter, a distinct signature of ``2 jets + $e^\pm$ 
+ $\mu^\mp$ + missing energy'' may be observed.  This same final state 
is also possible in the model of Ref.~\cite{m14} and has been analyzed in 
Ref.~\cite{mn14}, but the topology here is different and will be 
studied further elsewhere.

In conclusion, a syndetic model of fundamental interactions has been 
presented in the context of having one and only one Higgs boson in 
accordance with the standard model. The difference here is that 
radiative masses are obtained for the first family of quarks and leptons, 
as well as for all neutrinos.  The $U(1)_{PQ}$ symmetry is implemented 
to solve the strong $CP$ problem, in such a way that an exactly conserved 
residual $Z_2$ discrete symmetry remains to support a candidate particle 
for cold dark matter.  The same $Z_2$ symmetry enables the one-loop 
radiative masses.  The new particles required are possibly observable 
at the Large Hadron Collider in the near future.

\medskip
This work is supported in part 
by the U.~S.~Department of Energy under Grant No.~DE-SC0008541.

\bibliographystyle{unsrt}

\end{document}